\documentclass[final,3p,times, twocolumn]{elsarticle}

\usepackage{amssymb}
\usepackage{amsmath}
\usepackage[utf8]{inputenc}
\usepackage[T1,T2A]{fontenc} 
\usepackage{graphicx}
\usepackage{esvect}
\usepackage{placeins}
\usepackage{amsmath}
\usepackage{hyperref}
\usepackage{braket}
\usepackage[onehalfspacing]{setspace}
\usepackage{caption}
\usepackage{amsmath}
\usepackage{booktabs}
\usepackage{subfig}
\usepackage{float}
\usepackage{cleveref}
\usepackage[version=3]{mhchem}

\journal{Electrochimica Acta}

\begin{document}

\begin{frontmatter}



\title{The intrinsic electrostatic dielectric behaviour of graphite anodes in Li-ion batteries -- across the entire functional range of charge}


\author[inst1,inst2]{Simon Anniés}
\author[inst1]{Christoph Scheurer}
\author[inst1]{Chiara Panosetti}


\affiliation[inst1]{organization={Fritz-Haber-Institute of the Max-Planck-Society},
 addressline={Faradayweg 4-6}, 
 postcode={14195}, 
 city={Berlin},
 country={Germany}}
 
\affiliation[inst2]{organization={Chair for Theoretical Chemistry, TU Munich},
 addressline={Lichtenbergstr. 4}, 
 postcode={85747}, 
 city={Garching b. M\"unchen},
 country={Germany}}

\begin{abstract}
Lithium-graphite intercalation compounds (Li-GICs) are by far the most common anode material for modern Li-ion batteries. However, the dielectric response of this material in the electrostatic limit (and its variation depending on the state of charge (SOC)) has not been investigated to a satisfactory degree -- neither by means of theory nor by experiment -- and especially not for the higher range of SOC. Nevertheless, said dielectric behaviour is a highly desired property, particularly as an input parameter for charged kinetic Monte Carlo simulations -- one of the most promising modeling techniques for energy materials. In this work, we make use of our recently published DFTB parametrization for Li-GICs based on a machine-learned repulsive potential in order to overcome the computational hurdles of sampling the long-ranged Coulomb interactions within this material -- as experienced by the charge carriers within. This approach is rather novel due to computational cost, but best suited for investigating our specific property of interest. For the first time, we discover a mostly linear dependency of the relative permittivity $\epsilon_r$ on the SOC, from $\approx 7$ at SOC 0\% to $\approx 25$ at SOC 100\%. 
In doing so, we also present a straightforward approach that can be used in future research for other intercalation compounds -- once sufficiently fast and long-ranged computational methods -- such as linear-scaling DFT, a good DFTB parametrization, or atomic potentials with inbuilt electrostatics -- become available. However, while the presented qualitative behaviour is robust and our results compare favourably with the very few experimental studies available, we do stress that the quantitative results are strongly dependent on our estimation of the partial charge transfer from the intercalated Li-ion to the carbon host structure, and need to be verified by further experiments and other calculations. Yet, our research shows that in principle, two measurements -- one at low and one at high SOC -- should suffice for that purpose.
\end{abstract}


\begin{highlights}
\item The relative permittivity of lithium intercalated graphite anodes is linearly dependent on the state of charge, with approximate values of around 7 at 0\% and around 25 at 100\%.
\item Sampling the Coulomb interaction between two charge carriers within the material is enabled by our recently published DFTB parametrization for lithium intercalated graphite.
\item The dielectric screening in graphite is significantly larger in the direction perpendicular to the graphene-sheets, than in the direction parallel to them. This is important when comparing experimental results for graphite powder and for perfect graphite crystals.
\end{highlights}


\end{frontmatter}



\section{INTRODUCTION:}
\label{sec:intro}

The relative permittivity (RP) is one of the defining properties of many materials, as it describes the degree to which the Coulomb interactions between charge carriers are screened within the material compared to vacuum. Especially in older literature, this property is also known as ``dielectric constant'', even though it is far from being constant, but dependent on temperature, the frequency of probing electric fields and even the underlying mechanisms in terms of polarizability, conductivity and others.

Traditionally, the RP has primarily been of interest for insulators, but in recent times it has also been increasingly investigated for conducting materials~\cite{chung2021factors}, where it stems from the interaction between a small fraction of the charge carriers and the atoms. The fact that these charge carriers are mobile in conducting materials fundamentally changes the way the property and its dependencies need to be understood within these materials, compared with insulators.

\begin{figure}[t]
 \centering
 \includegraphics[width=0.99\columnwidth]{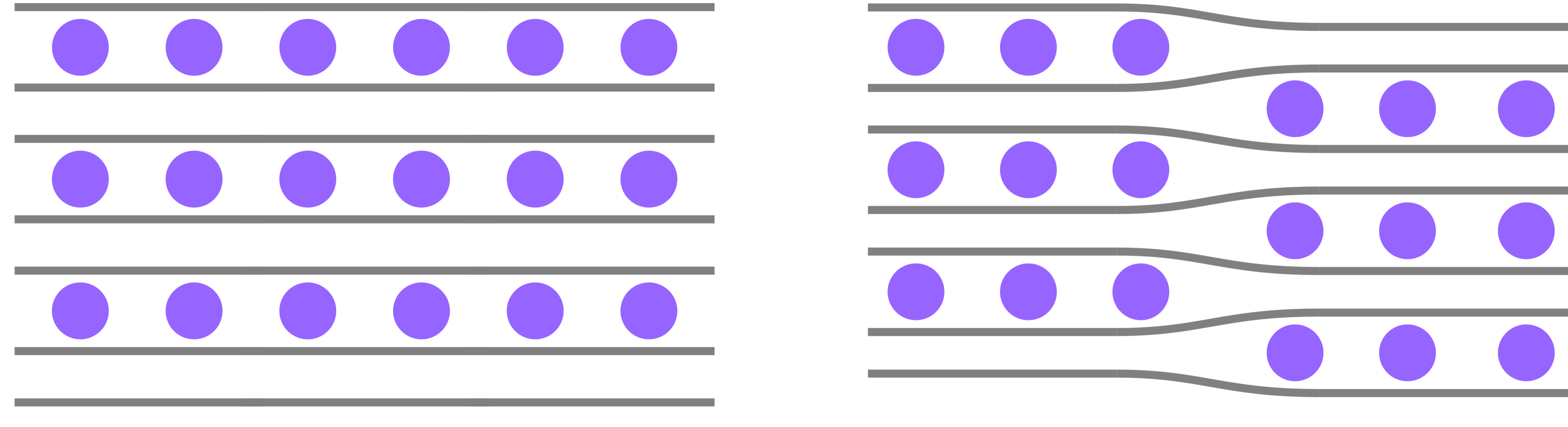}
 \caption{Illustration of lithium-ions (purple spheres) intercalated into a graphite host structure (grey lines) at $50\%$ state of charge, based on the Daumas-Herold domain model~\cite{Daumas1969}. Li-ions tend to fill up every second layer completely (staging, left), before starting to intercalate into the other half of the layers. However, this behaviour is not global, but occurs in finite-sized domains (right) and is not expected to be perfect in real systems. During the intercalation process, the distance between the graphene sheets (interlayer distance) is increased by around $10\%$.}
 \label{fig:staging}
 \end{figure}

One group of materials of great interest are lithium-graphite intercalation compounds (Li-GICs), which constitute by far the most common anode material in modern lithium ion batteries. During the charging and discharging cycles of the battery, Li-ions are stored between the layers of the graphite host structures, up to a stoichiometry of \ce{LiC6}, which traditionally translates to a state of charge (SOC) of $100\%$ and corresponds to one Li-ion above every third \ce{C6} ring of the hexagonal base lattice -- even though recent studies have shown that overlithiation beyond that point is possible at ambient conditions~\cite{grosu2021lithium}. The distribution of the Li-ions for intermediate SOCs is not uniform, but ordered as shown in~\cref{fig:staging}, as first explained in~\cite{Daumas1969}.
 
Within the scope of this specific material, we define the RP we are investigating as ``the damping of the electrostatic interaction between two Li-ions or Li-ion vacancies embedded in the material caused by the surrounding charge-carrier density''. As of note, this is the \textit{electrostatic}, low-frequency RP, as opposed to what is measured in many experiments, which make use of alternating AC-fields at a vast variety of frequencies~\cite{chung2021factors, hotta2011complex}. Furthermore, we point out that this specific property is directionally separated -- its contribution in the $xy$-plane (in this work defined as parallel to the graphene sheets) is expected to be different from the contribution in $z$-direction (orthogonal to the graphene sheets). This makes direct comparison with experiments performed on graphite powder as opposed to a ``perfect'' crystal non-trivial -- a problem we address in~\cref{sub:z_dir}.
 
One of the primary motivations for investigating the RP of Li-GICs is the fact that it is a required input parameter for including charge in kinetic Monte Carlo simulations (kMCs), which in turn are a crucial method for studying charge carrier dynamics in functional energy materials~\cite{gavilan2020kinetic, gavilan2021kinetic, methekar2011kinetic, dean2021overscreening}. Long-ranged Coulomb interactions are a necessity when performing kMC on systems, which include charged or partially charged particles. For example, Casalegno et al.~\cite{casalegno2010methodological} have shown that not including such interactions (as it would be the case when using e.g. force field approaches with finite-size descriptors) causes an error of 14\% in the protonic diffusion coefficients in doped perovskites. 
The situation becomes significantly more complicated when looking at anode and cathode materials, but also electrolytes, perovskites in solar cells, and any other type of functional materials that involves changes in the density and/or local ordering of charge carriers as part of their intended function. This is due to the fact that the local relative permittivity then is not only influenced by the ``host'' material, but also by the charge carriers close by. Therefore, the RP of Li-GICs changes significantly depending on the SOC ({\em vide infra}).

In this work, we put forward a systematic approach to investigating this crucial property. Based on this, we determine for the first the relative permittivity of Li-GICs, as a function of the SOC, for the entire functional range of the material.
 
Beyond the previously outlined interest for charge-kMC, we believe there to be many more valuable applications for the relative permittivity of Li-GICs also at higher levels of the multiscale simulation hierarchy: one such motivation is understanding charge gradients, as they occur during the fast charging of modern batteries in electric vehicles, and the chemical pressure which leads to plating and dendrite formation inside the batteries under certain operational conditions. The latter phenomenon is typically investigated by means of continuum simulations like e.g. by Hein et al.~\cite{hein2020electrochemical}, which also rely at least implicitly on knowledge of the dielectric response. Furthermore, a simple model of the charge carrier electrostatics could be used as a physical baseline for otherwise short-ranged machine learning models or cluster expansions. Another related field is the development of functional materials based on doped graphite~\cite{reed2010effective}. 

The dielectric behaviour of Li-GICs (and solid materials in general) is significantly more complicated than the expression ``dielectric constant'' would suggest, and is governed by vastly different physics at different frequencies of a probing field. In the static limit (which this work aims to investigate -- the probing field is essentially the electrostatic field of the intercalated ion itself), no periodic movement (beyond thermal fluctuations) of the electrons is induced. However, this picture changes in the kHZ range, where the entire charge carrier density oscillates with the probing field, causing large polarization and large dielectric screening. For example, Chung et al.~\cite{chung2021factors} measured an RP of $\epsilon_r = 2100$ for highly oriented pyrolytic graphite (HOPG) and even higher ones for other carbon structures, at 2-10KHz. Moving on to the GHz regime, a balance is reached where the field oscillations are too fast for macroscopic bulk currents to build up, and a situation occurs that is arguably similar to the static limit and \textit{may} serve as comparison for our research. Hotta et al.~\cite{hotta2011complex} put forward a dielectric constant for graphite powder of $\epsilon_r \approx 15$, at 6GHz. Finally, at even higher frequencies beyond THz, the electric field becomes high enough in energy to excite a significant number of electrons, again creating a physically different situation with much lower dielectric screening, which converges to transparency in the $\omega \rightarrow \infty$ limit. A study by Jellison et al.~\cite{jellison2007measurement} in the frequency regime of visible light finds an RP in $xy$-plane for HOPG of $\epsilon_r = 4.21$, which, for the previously mentioned reasons, cannot be used as comparison either and is expected to serve as a lower bound in the following.
 
It is apparent that there is a glaring lack of studies investigating the \textit{exact} property of interest to us, which is -- again -- the electrostatic dielectric response of a perfect graphite crystal in $xy$-plane, i.e. parallel to the graphene sheets, as experienced by some internal charge carriers (in this case Li-ions and vacancies). There are some studies available on graphene, either on some substrate or quasi-freestanding, with results ranging from $\epsilon_r = 2.2-5.0$ by Elias et al.~\cite{elias2011dirac} to $\epsilon_r = 15.4$ by Reed et al.~\cite{reed2010effective}, and another study by Bostwick et al. finding $\epsilon_r \approx 4.4$~\cite{bostwick2010observation}, none of which can serve as direct comparison to our research either. However, there is a study on bilayer graphene (which according to our calculations can be compared with graphite quite well) by Bessler et al.~\cite{bessler2019dielectric}, putting forward an RP of $\epsilon_r = 6\pm 2$. This is likely the most reliable direct experimental comparison currently available to us.

In terms of theoretical approaches to determining the dielectric response of materials, substantial work has been done on water~\cite{raabe2011molecular, aragones2011dielectric, sharma2007dipolar, ruiz2018quest}. There is also some promising work by Gigli et al.~\cite{gigli2021thermodynamics} in the development of an integrated machine learning model predicting the dielectric response of $BaTiO_3$. However, all these methods are reliant on the presence of polarizable dipoles within the system, which is not the case for Li-GICs.

\section{METHODOLOGY:}
\label{sec:methodology}
\subsection{Computational details:}
\label{sec:comp}

For this study, we used self-consistent-charge Density Functional Tight Binding (SCC-DFTB~\cite{elstner1998self}) as implemented in in DFTB+~\cite{Hourahine2020}, with the parametrization developed in our group. The corresponding Slater-Koster files are available upon request and have been parametrized and tested as described in~\cite{panosetti2021dftb, annies2021accessing}. The repulsion potential is machine-learned by means of Gaussian Process Regression (GPR)~\cite{panosetti2020learning}.

Geometries have been constructed and analysed by means of the Atomic Simulation Environment (ASE~\cite{Bahn2002}) which we also used as a base framework for all force- and energy-calculations, structure relaxations (specifically using the Broyden-Fletcher-Goldfarb-Shanno (BFGS) algorithm as an optimizer~\cite{Shanno1985}), and transition state calculations. For the latter, we employed the Nudged Elastic Band (NEB,~\cite{berne1998classical}) algorithm with the BFGS optimizer and climbing image~\cite{Henkelman2000,Henkelman2000a} switched on.

For all DFTB calculations, we used a well converged k-point density of at least 0.1/\r{A} for the $z$-component of the unit cell. The $xy$ size of the cell is large enough to sample at the Gamma point. The SCC-tolerance is $10^{-6}$. We employed Fermi filling with a Fermi temperature of 0.001 Kelvin, as well as a Broyden mixer~\cite{johnson1988modified} for convergence acceleration with a mixing parameter of 0.5. All of these settings have been tested with regard to convergence for the whole range of SOC. As described in~\cite{panosetti2021dftb}, our parametrization is meant to be used with the Lennard-Jones dispersion correction~\cite{zhechkov2005efficient} switched on.

In terms of supercell size convergence, satisfactory convergence of the extracted slopes of the energies relative to the inverse distances (see~\cref{sec:methodology} for details) is reached at distances of around $15\text{\r{A}}$ between the periodic images of the sampled areas of the investigated layers (see~\cref{fig:inv_layer_Li} and~\cref{fig:inv_layer_vac}), a value we reach or exceed with all used supercells. We point out that convergence of total energies is still not reached at those distances due to the long-ranged nature of the Coulomb interactions, however our property of interest -- the previously mentioned slope -- is rather robust to total energy shifts of the whole sampled area. Slight further improvements would still be probable with even larger supercells, but due to computation time constraints, and due to the fact that other effects introduce much larger errors to the final results, we chose the supercells described in the following sections.

\subsection{Derivation of the relative permittivity $\epsilon_r$ from the Coulomb law and discussion of the partial charge transfer:}

In order to determine the relative permittivity within the system, our approach is to sample the electrostatic interactions between two charge carriers inside the system, placed at varying distances from each other. We begin our considerations with the electrostatic energy $E_{Coul}$ of two charge densities $\rho_1$ and $\rho_2$, governed by the Coulomb law

 \begin{align} E_{Coul} = \frac{1}{4\pi \epsilon_0 \epsilon_r} \cdot \int \int \frac{\rho_1 \cdot \rho_2}{|\vec{r}_1 - \vec{r}_2|} dr_1 dr_2 \label{Diel:eq:1} \end{align}
 
where $\epsilon_0$ is the vacuum permittivity. In this work, we chose to approximate the sampled charge carriers as point charges, an approximation that is unproblematic for Li-ions, but requires a bit more consideration for Li-ion vacancies, which may be more diffuse in shape. In this approximation, the expression becomes
 \begin{align} E_{Coul} = \frac{1}{4\pi \epsilon_0 } \cdot \frac{Q_1 \cdot Q_2}{\epsilon_r} \cdot \frac{1}{|\vec{r}_1 - \vec{r}_2|} \label{Diel:eq:2} \end{align} 

which has been rearranged to clearly separate the continuum electrostatic from the geometric quantities.
With the two point charges being Li-ions or Li-ion vacancies intercalated into the graphite host structure, we get
 
 \begin{align} E_{Coul} = \frac{e^2}{4\pi \epsilon_0} \cdot \frac{Z^2_{Li}}{ \epsilon_r} \cdot \frac{1}{R} \label{Diel:eq:3} \end{align} 

where $e$ is the electron charge, $Z_{Li}$ the partial charge on the Li-ion and $R$ the distance between the two charge carriers. In order to link this Coulomb energy to the potential energy $E_{pot}$ of an entire supercell (as we obtain from our DFTB calculations), an appropriate reference energy $E_0$ needs to be introduced, which is the host structure energy in the limit of $R \rightarrow \infty$ at that specific stoichiometry. With this we finally get:

 \begin{align} E_{pot} = \frac{e^2}{4\pi \epsilon_0} \cdot \frac{Z^2_{Li}}{ \epsilon_r} \cdot \frac{1}{R} + E_0 \label{Diel:eq:4} \end{align} 

The first two terms of this expression can be accessed as the slope of a linear regression, when plotting $E_{pot}$ over $1/R$. In order to relate said slope to $\epsilon_r$, we approximate $E_{Coul} \approx E_{pot} - E_0$. While this obviously holds in the macroscopic limit for a homogeneous medium, here we are attempting to measure the Coulombic repulsion felt by two particles on a potential energy surface (PES) resolved at the atomistic level. The latter is not infinitely smooth, but presents minima, maxima and saddle points due to the carbon host structure. It however shows a sufficiently regular pattern (as shown in~\cite{annies2021accessing}) to assume that analogous points (i.e., comparing minima with minima etc.) exhibit analogous local shape. Consistently, the DFTB total energy is indeed expressed (within the formalism of the method \cite{koskinen2009density}) as a sum of three contributions: the so-called band structure energy, the Coulomb energy (the expression of which depends explicitly on the partial charges on the individual atoms, {\em vide infra}), and the repulsive energy. Ideally, we can isolate the Coulomb energy between two particles in the system by considering ``the rest of the DFTB total energy'' as a background to subtract pointwise, assuming that it will be similar enough in similar local environments. Ideally, this would correspond to the limit for infinitely dilute, unperturbed PES, where the Coulombic interaction between charge carriers vanishes. In this light, for each sampled point we may identify a suitable (pointwise) reference $E_0$ as the corresponding point on such unperturbed PES, e.g., in the case of minima, the total energy of a minimum at the center of any \ce{C6} ring of the graphitic host sufficiently far away from the other charge carrier.

We point out that this approximation neglects distortions of the local structure and electron density that may be caused by two charge carriers being close together -- an approximation that holds well for the low SOC regime, but not quite as well for the high SOC regime, as will be shown. The reference energy $E_0$ does not impact that slope, but only causes an up or down shift along the $y$-axis. Therefore, it is not immediately relevant for the extraction of the $\epsilon_r$, but it is needed for consistent plotting of the Coulomb energy. For this reason, we estimate $E_0$ by interpolating the slope to the $1/R \rightarrow 0$ limit (see the following chapters for more details).

The only other variable that is left, then, is the partial charge of the intercalated lithium $Z_{Li}$. It is known that, upon intercalation, the electron density of the Li-atom is partially transferred to the carbon host structure, leaving the intercalant as something usually denominated as $Li^+$, but the exact magnitude of said charge transfer is hard to pinpoint -- and not an observable. The self-consistent-charge cycles of our DFTB+ calculations output a local electron population of around $0.21$ to $0.26$ for the Li-intercalants depending on the local environment of the respective Li-ion, which corresponds to a partial charge of $+0.79e$ to $+0.74e$. However, literature reports a variety of different values. Valencia et al.~\cite{valencia2006lithium} pointed out the large dependency of the partial charge transfer on the method of analysis and presented values of $0.43e$ (Mulliken charge analysis), $0.47e$ (Voronoy), $0.6e$ (Löwdin) and $1.0e$ (Bader). Krishnan et al.~\cite{krishnan2013revisiting} found $0.86e$, also by Bader analysis. Song et al.~\cite{song2001structure} determined $0.68e$ by means of quantum mechanical calculations and comparison with experimental layer spacing, but also pointed out that this value may change with SOC. Finally, Rakotomahevitra et al.~\cite{rakotomahevitra1992electronic} calculated $0.517e$, using an extra-orbital model. To complicate things further, $\epsilon_r$ depends on $Z_{Li}$ quadratically (see~\cref{Diel:eq:4}) and is therefore very sensitive to it. For consistency, we will move forward in this work assuming $Z_{Li}=0.765\pm 0.05$ -- the median partial charge to which the SCC converges. The latter indeed directly enters the expression for the Coulombic contribution to the DFTB total energy, as mentioned before, and therefore represents a natural choice within the framework (and within the approximations) of DFTB. 
We add a generous, but arbitrary measure of uncertainty, and we stress again that all absolute numbers for the relative permittivity $\epsilon_r$ presented in this work need to be understood with this assumption in mind. However, the same is not true for the electrostatic screening (captured by the slopes of the linear regression) -- this can be taken at face value and may be, depending on the model or method employed, the more valuable property to take into account for future research.
 
\section{RESULTS \& DISCUSSION:}
\label{sec:results_discussion}

\subsection{Lithium-intercalant pairwise interaction screening ($xy$-plane):}
\label{sub:li_xy}

\begin{figure}[h]
 \centering
 \includegraphics[width=0.99\columnwidth]{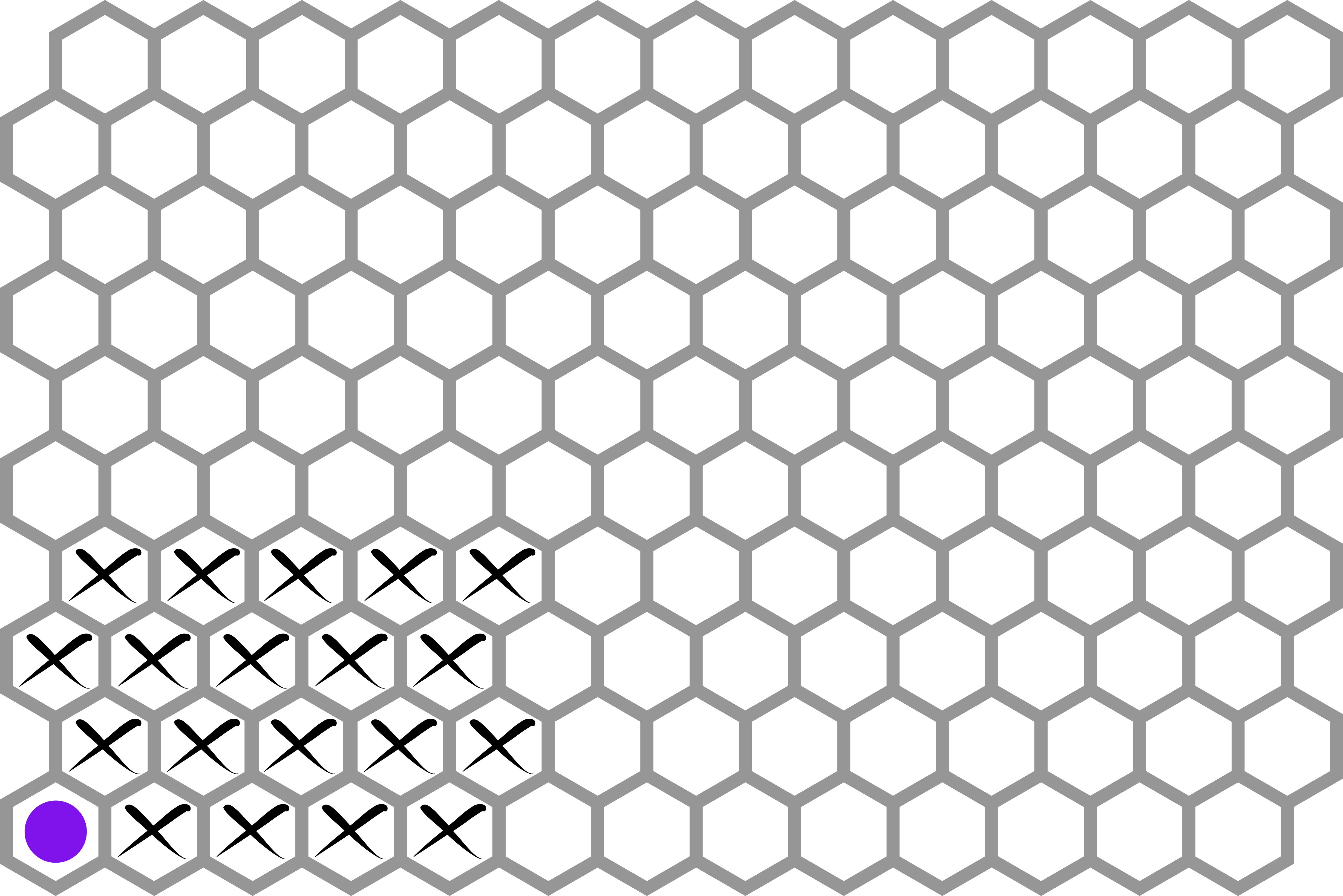} 
 \caption{Illustration of the investigated layer for pairwise Coulombic Li-ion-interactions with one fixed Li-ion ($purple$). Of the locations marked with an ``X'', one is occupied by, the other Li-ion, while all others are empty. This layer will be referred to as $inv(Li)$ throughout this work.}
 \label{fig:inv_layer_Li}
 \end{figure}

In order to sample the pairwise Li-ion interactions within the material, we construct an investigated layer (called $inv(Li)$, consisting of 300 carbon atoms and two Li-ions, one of them fixed in the corner, the other one sampling a number of different positions, as schematically illustrated in~\cref{fig:inv_layer_Li}. We then combine this investigated layer with varying stackings of empty and filled layers (as they predominantly appear in the system according to the staging model), perform full structure relaxations on the resulting supercells (with two or three layers and 600-900 carbon atoms) for each of the 21 possible Li-ion positions (per data point on the SOC axis) and extract the potential energies of the cells. We also perform nudged elastic band (NEB) calculations between all combinations of two relaxed structures, which have directly adjacent occupied Li-positions, thus acquiring 41 transition states (per data point) of diffusive next-neighbour jumps. This is especially relevant for kinetic Monte Carlo applications, where the dynamics is governed by the rates (and therefore, the activation barriers) corresponding to elementary processes bringing the system from one state to another. As our results show, the Coulombic behaviour we get is identical (within method accuracy) for both cases (\cref{fig:inv_layer_Li}). This justifies the application of simplified rescaling rules solely based on Coulombic interactions to include the effect of next-neighbour occupations on the elementary barriers, allowing to set up a charged kMC model with one, simple, elementary barrier (the jump of one Li-ion from one site to another in the infinitely dilute limit), and correcting the latter depending on the number and distance of nearest neighbours as well as on the direction of the jump.

\begin{figure*}[t]
\includegraphics[width=0.99\columnwidth]{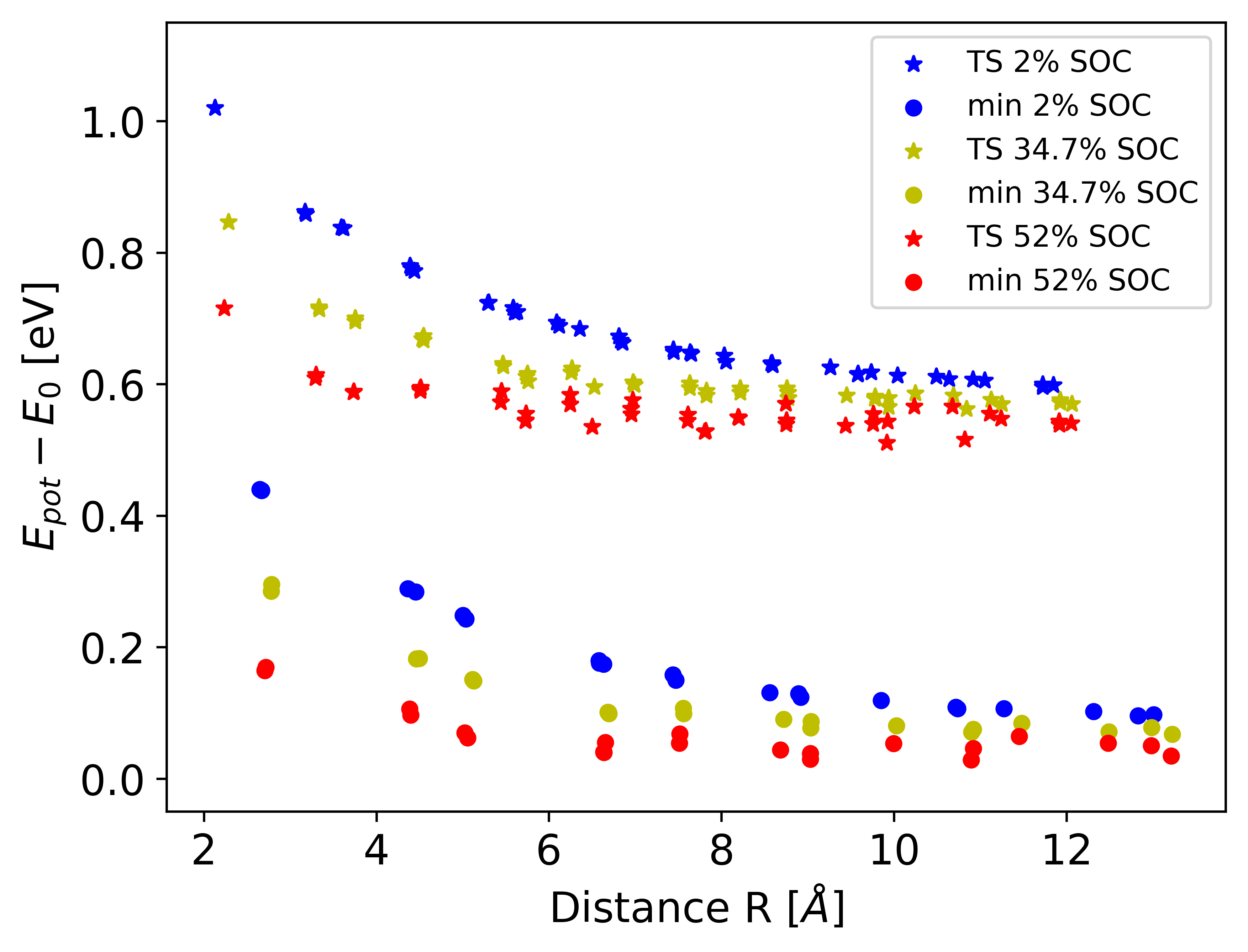} 
\includegraphics[width=0.99\columnwidth]{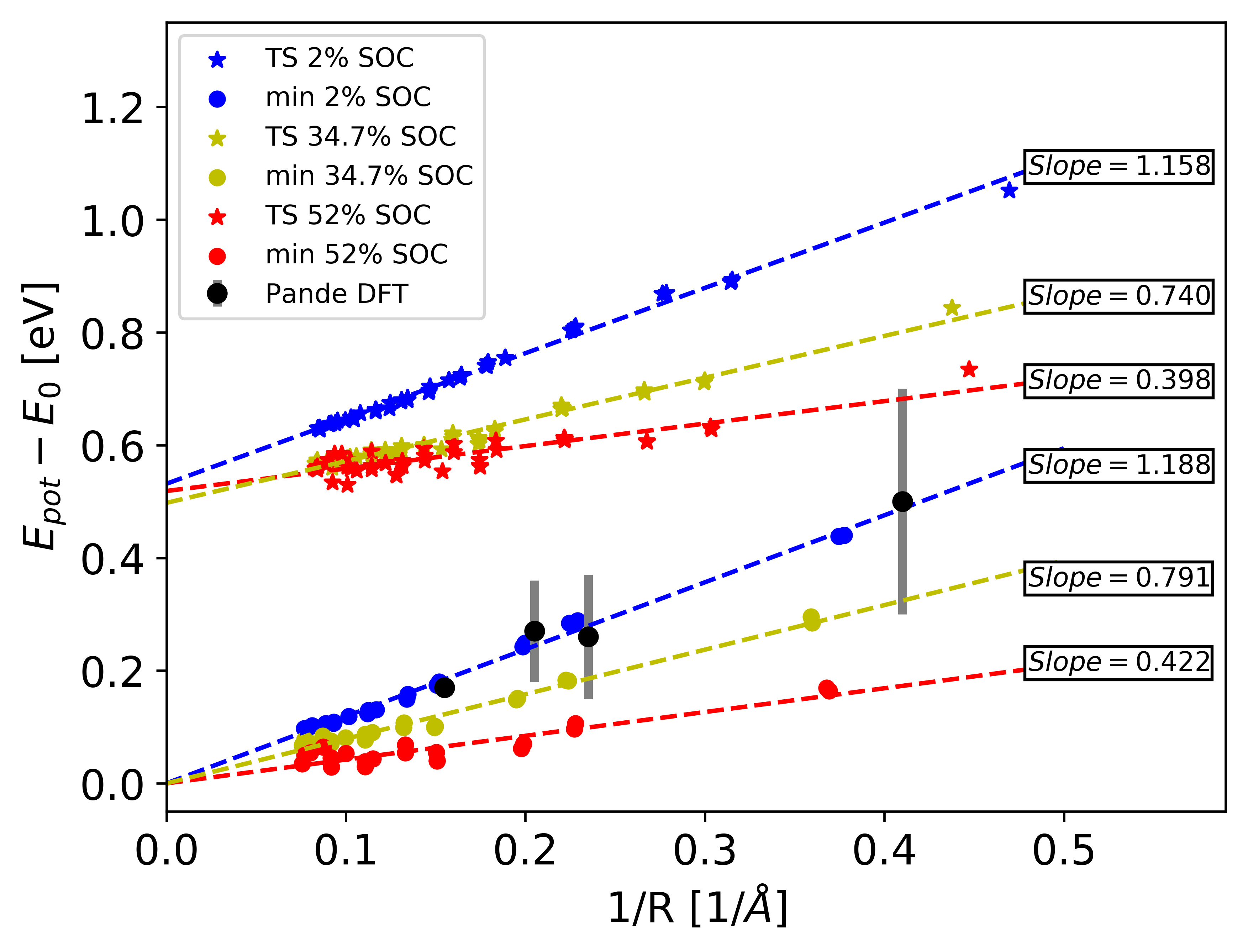} 
\caption{Approximate Coulomb energies $E_{Coul} \approx E_{pot} - E_0$ of minima and transition states relative to $E_0$ (at the respective SOC) in dependence of the distance (\textit{left}) and inverse distance (\textit{right}) between the investigated Li-ions at different SOCs. The slopes have been extracted by means of linear regression. The global minima $E_0$ in the $1/R \rightarrow 0$ limit are extrapolated from the linear regression of the slopes of the minima and set to $0$ in the plot. The DFT reference is taken from an Ising model based on BEEF-vdW DFT~\cite{pande2018robust}. The data points represent the pair-interactions for next-neighbour, second, third and fourth next neighbour lithium positions. The SOC is not clearly defined without giving the size of the otherwise empty supercell (a limitation of the Ising model), but can be understood as ``low'' and taken as comparison for our blue data-points.}
\label{fig:results_Li}
\end{figure*}

 \begin{table*}[h] 
 \centering
 \begin{tabular}{lllll}
 \toprule
 \textbf{Structure} & \textbf{Stoichiometry} & \textbf{SOC} & \textbf{Slope [eV\text{\r{A}}]} & \textbf{rel. permittivity}\\
 \midrule
 $inv(Li) - empty$ & \ce{Li2C_{600}} & 2.0\% & $1.168\pm 0.013$ & $7.23\pm 0.86$ \\ 
 $bilayer\ graphene$ & \ce{Li2C_{600}} & 2.0\% & $1.111\pm 0.018$ & $7.55\pm 1.04$ \\ 
 $inv(Li) - empty - full$ & \ce{Li_{52}C_{900}} & 34.7\% & $0.759\pm 0.018$ & $11.32\pm 1.74$ \\
 $inv(Li) - full$ & \ce{Li_{52}C_{600}} & 52.0\% & $0.407\pm 0.026$ & $21.04\pm 4.05$ \\
 \bottomrule
 \end{tabular}
 \caption{Results for the data points constructed with an otherwise empty investigated layer with sampled lithium positions \textit{inv(Li)}. The slopes and their RMSEs stem from linear regression of all data points (minima and transition states) -- shifted to the same baseline. The corresponding relative permittivities have been calculated via the Coulomb law in~\cref{Diel:eq:4}.}
 \label{tab_inv_Li}
 \end{table*}
 
By plotting the estimated Coulomb energies $E_{Coul} = E_{pot} - E_0$ as functions of the (inverse) distance between the two Li-ions in the investigated layer, we clearly illustrate how our model captures the Coulombic nature of the interaction close to perfectly in the case where no other Li-ions are present in the adjacent layers (\cref{fig:results_Li}, blue) that could distort the electron density. The very minor scatter in this case likely stems from the small distortions in the carbon structure close to the Li-intercalants and the fact that our full cell relaxations cannot be converged to infinity, but have to be stopped at some threshold forces. In the other two cases, there is some more scatter present (caused by the slightly deformed charge density, due to the filled adjacent layers (one in case of SOC 34.7\%, two in case of SOC 52\%)) but the overall behaviour is still predominantly Coulombic.
 
In a next step, we extract the slopes from the $1/R$-plots by means of linear regression (see~\cref{tab_inv_Li}). We achieve this by shifting the transition state energy levels down to the ground states and then fit all data points at once. Based on that, the relative permittivity $\epsilon_r$ can be extracted from the Coulomb law in~\cref{Diel:eq:4}. For the sake of comparison with experiment, we also perform the same procedure with freestanding bilayer graphene, consisting of one $inv(Li)$ layer between 2 graphene sheets, which according to our model, has a very similar $\epsilon_r$ as (periodic) graphite at the same stoichiometry.

As pointed out in~\cref{fig:results_Li}, there is one DFT study available for comparison by Pande et al.~\cite{pande2018robust}, and the slope they find at low SOC agrees well with ours. However, due to the computational cost of DFT, they were only able to provide 4 data points, and only at the low end of the range of charge, which is the cheapest to compute.

Furthermore, Bessler et al. measured a relative permittivity of $\epsilon_r = 6\pm 2$ for bilayer graphene, which agrees quite well with our result of $\epsilon_r = 7.55\pm 1.04$.
	
\subsection{Vacancy pairwise interaction screening ($xy$-plane):}
\label{sub:vac_xy}

\begin{figure}[t]
 \centering
 \includegraphics[width=0.99\columnwidth]{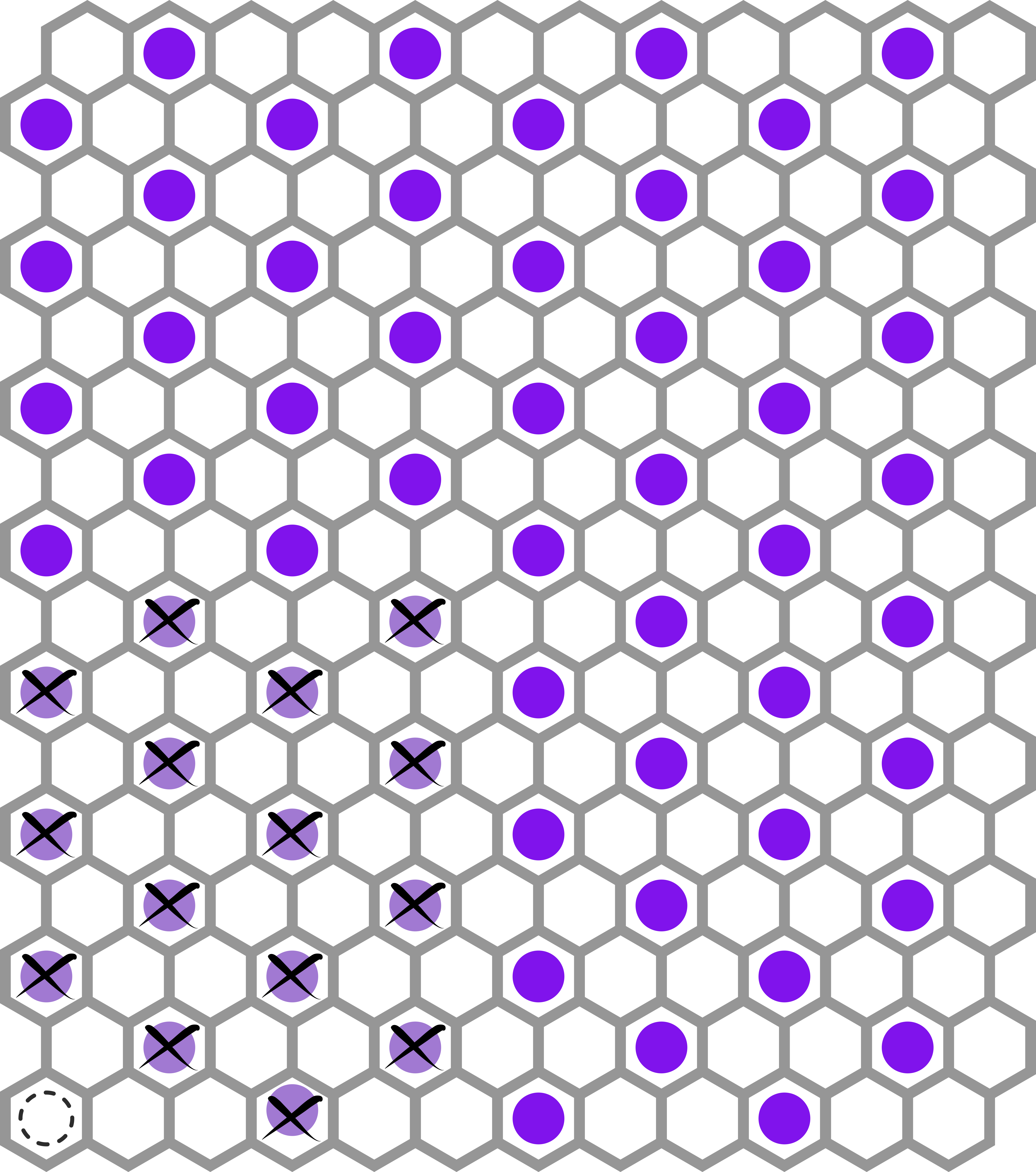}
 \caption{Illustration of the investigated layer for pairwise Coulombic vacancy-interactions with fixed Li-ions (\textit{purple}) and fixed vacancy \#1 (bottom left, dashed circle). Of the positions marked with an ``X'', the sampled vacancy \#2 is located on one, while all others are occupied by Li-ions. This layer will be referred to as $inv(Vac)$ throughout this work.}
 \label{fig:inv_layer_vac}
 \end{figure}

It is clear that it is not possible to investigate SOCs close to 100\% following the same approach, i.e. using the same investigated layer described in the previous chapter. Furthermore, for the higher SOCs, the diffusion mechanism transitions towards vacancy hopping instead of Li-intercalant hopping. Therefore, we introduce a second type of investigated layer (\cref{fig:inv_layer_vac}), which samples the pairwise Coulomb interactions between Li-vacancies within a filled layer instead (called $inv(Vac)$). A key difference here is the fact that, in a filled layer, only every third \ce{C6} ring is occupied by a Li-ion, so the investigated vacancies cannot be placed on each $C_6$-ring, but only on every third one. Due to this, a slightly larger cell is needed (384 carbon atoms) in order to sample a reasonable number of data points (15 per SOC), while also keeping the separation from the periodic image large enough to be well converged.

Another consequence of this is the fact that the diffusion path from one vacancy location to another is not clearly defined -- it may be a straight line or pass through one of the two next-neighbour minima in between. Because of this and because we already showed before that the slopes extracted from the minima and from the diffusion path intermediates are very similar, we limit our calculations to the minima in this chapter. Other than that, the procedure is the same as in~\cref{sub:li_xy}.

\begin{figure*}[t]
\includegraphics[width=0.99\columnwidth]{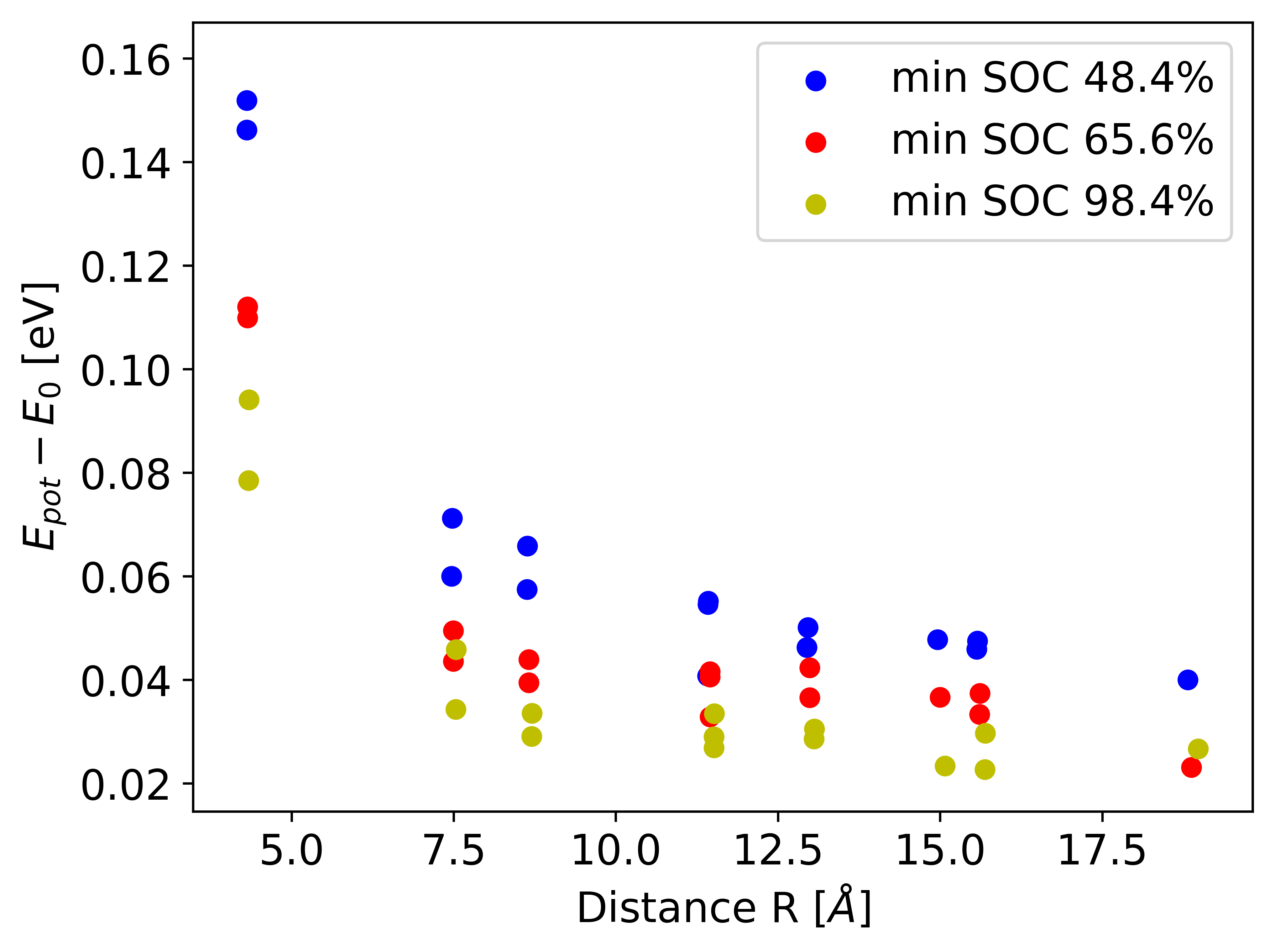}
\includegraphics[width=0.99\columnwidth]{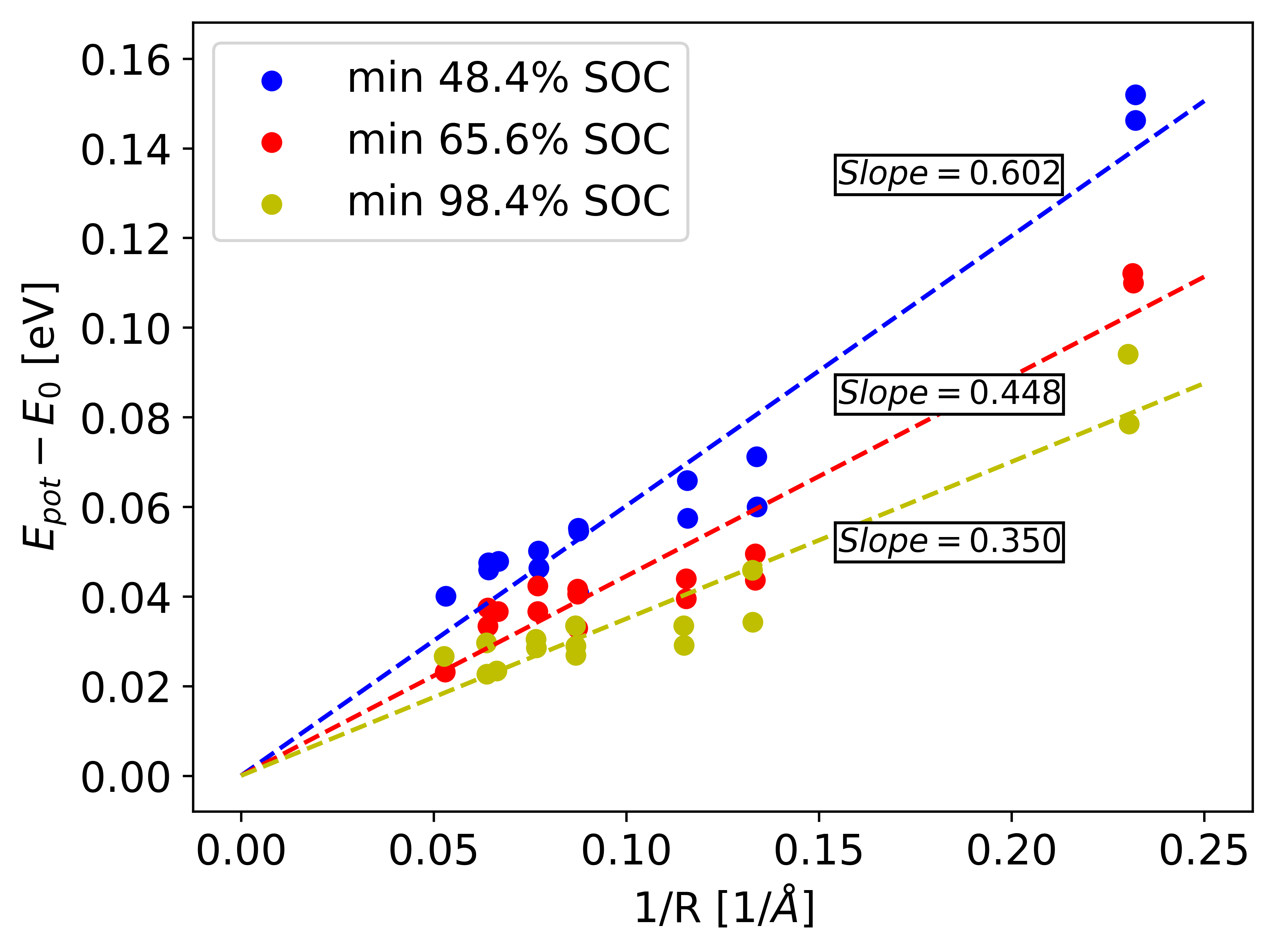}
 \caption{Approximate Coulomb energies $E_{Coul} \approx E_{pot}- E_0$ of local minima relative to $E_0$ (at the respective SOC) in dependence of the distance (\textit{left}) and inverse distance (\textit{right}) between the investigated Li-ion vacancies at different SOCs. Slopes have been extracted by means of linear regression. The global minima $E_0$ in the $1/R \rightarrow 0$ limit are extrapolated from the linear regression of the slope and set to $0$ in the plot.}
\label{fig:results_Vac}
\end{figure*}

 \begin{table*}[t] 
 \centering
 \begin{tabular}{lllll}
 \toprule
 \textbf{Structure} & \textbf{Stoichiometry} & \textbf{SOC} & \textbf{Slope [eV\text{\r{A}}]} & \textbf{rel. permittivity}\\
 \midrule
 $inv(Vac) - empty$ & \ce{Li_{62}C_{768}} & 48.4\% & $0.602\pm 0.047$ & $14.28\pm 2.95$ \\ 
 $inv(Vac) - empty - full$ & \ce{Li_{126}C_{1152}} & 65.6\% & $0.448\pm 0.041$ & $19.28\pm 4.22$ \\
 $inv(Vac) - full$ & \ce{Li_{126}C_{768}} & 98.4\% & $0.350\pm 0.034$ & $24.70\pm 5.55$ \\
 \bottomrule
 \end{tabular}
 \caption{Results for the data points constructed with a filled investigated layer with sampled vacancy positions \textit{inv(Vac)}. The slopes and their RMSEs stem from linear regression of the data points (minima). The corresponding relative permittivities have been calculated via the Coulomb law in~\cref{Diel:eq:4}.}
 \label{tab_inv_Vac}
 \end{table*}

 As can be seen in~\cref{fig:results_Vac}, there is more distortion to the Coulombic behaviour present in this case, some of which is even qualitative and cannot just be attributed to the uneven electron density in the adjacent layers. Especially, the data points at ``medium'' distance ($\approx 8\text{\r{A}}$ in the left plot or $\approx 0.125\ 1/\text{\r{A}}$ in the right plot of~\cref{fig:results_Vac}) have lower energy than expected if it was purely governed by the Coulomb law. As pointed out before, this is to be expected within our approximations, since a combination of multiple effects are at play here, that can not be addressed without explicit treatment of the overlapping local environments of the charge carriers: firstly, for these compositions, there is not just vacuum present in the space between the investigated sites (as it is the case in all configurations in~\cref{sub:li_xy}, as well as the closest possible vacancy positions (see~\cref{fig:inv_layer_vac}). This additional charge density between the sampled sites seems to have an additional stabilizing effect in our DFTB calculations. For vacancy pairs at distances larger than $\approx 10\text{\r{A}}$, this effect probably becomes negligible again, since there is hardly any interaction energy left to be screened at those distances. Secondly, the point-charge approximation may be less accurate for a Li-vacancy, than it is for a Li-ion, since the next-neighbour local minima within the host structure are not occupied either. And thirdly, upon inspection of the fully relaxed structures, one realizes that the Li-ions adjacent to the investigated vacancies are not located in the middle of their respective \ce{C6} ring, but slightly displaced towards the vacancy. Due to these additional effects, fitting the potential energies found for these configurations to the unperturbed Coulomb law ($E_{Coul} = E_{pot} - E_0$) is a more severe approximation in the context of the $inv(Vac)$ investigated layer than it was for the $inv(Li)$ investigated layer. Finally, the slopes and relative permittivities $\epsilon_r$ are summarized in~\cref{tab_inv_Vac}, analogous to~\cref{sub:li_xy}.
 
\subsection{Dilute adjacent layers:}
\label{sub:dilute}
 
One of the complications when investigating Li-GICs is the fact that the same stoichiometry can be realized in a variety of different ways, and while it is known that staged configurations are favoured in equilibrium and in perfect crystals, other -- dilute -- configurations may still play a role, when the system is under the effects of defects, grain boundaries or nonequilibrium states caused by fast charging. Because of this, we choose to investigate some of these configurations, as well. For this purpose, we make use of the same investigated layers as before, but combine them with an adjacent layer, that is dilutely filled at 50\% capacity. In order to realize the periodicity of said layer, a slightly larger cell with 360 carbon atoms is necessary in the case of the Li-ion investigated layer.
 
\begin{figure*}[t]
\includegraphics[width=0.99\columnwidth]{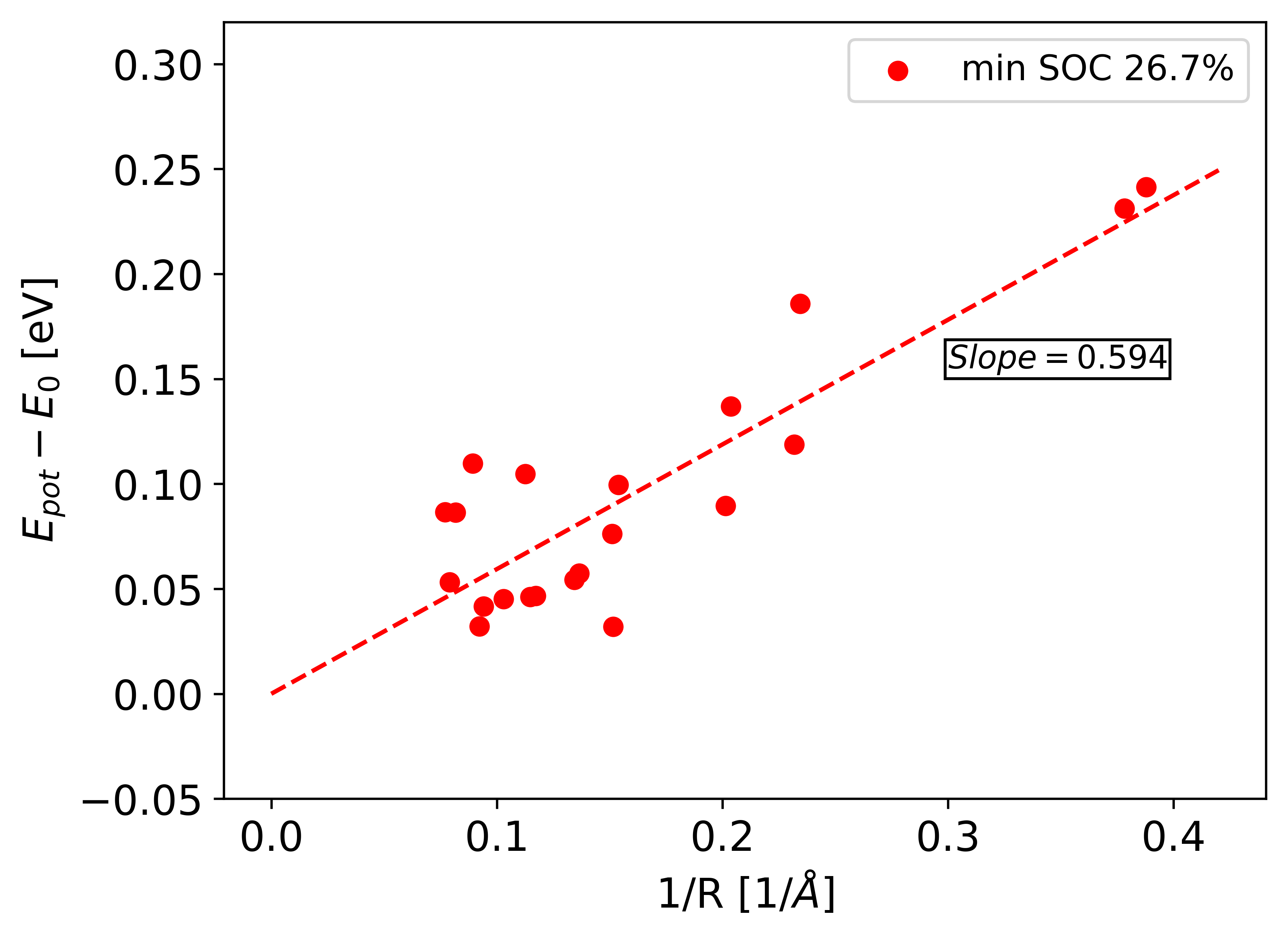} \includegraphics[width=0.99\columnwidth]{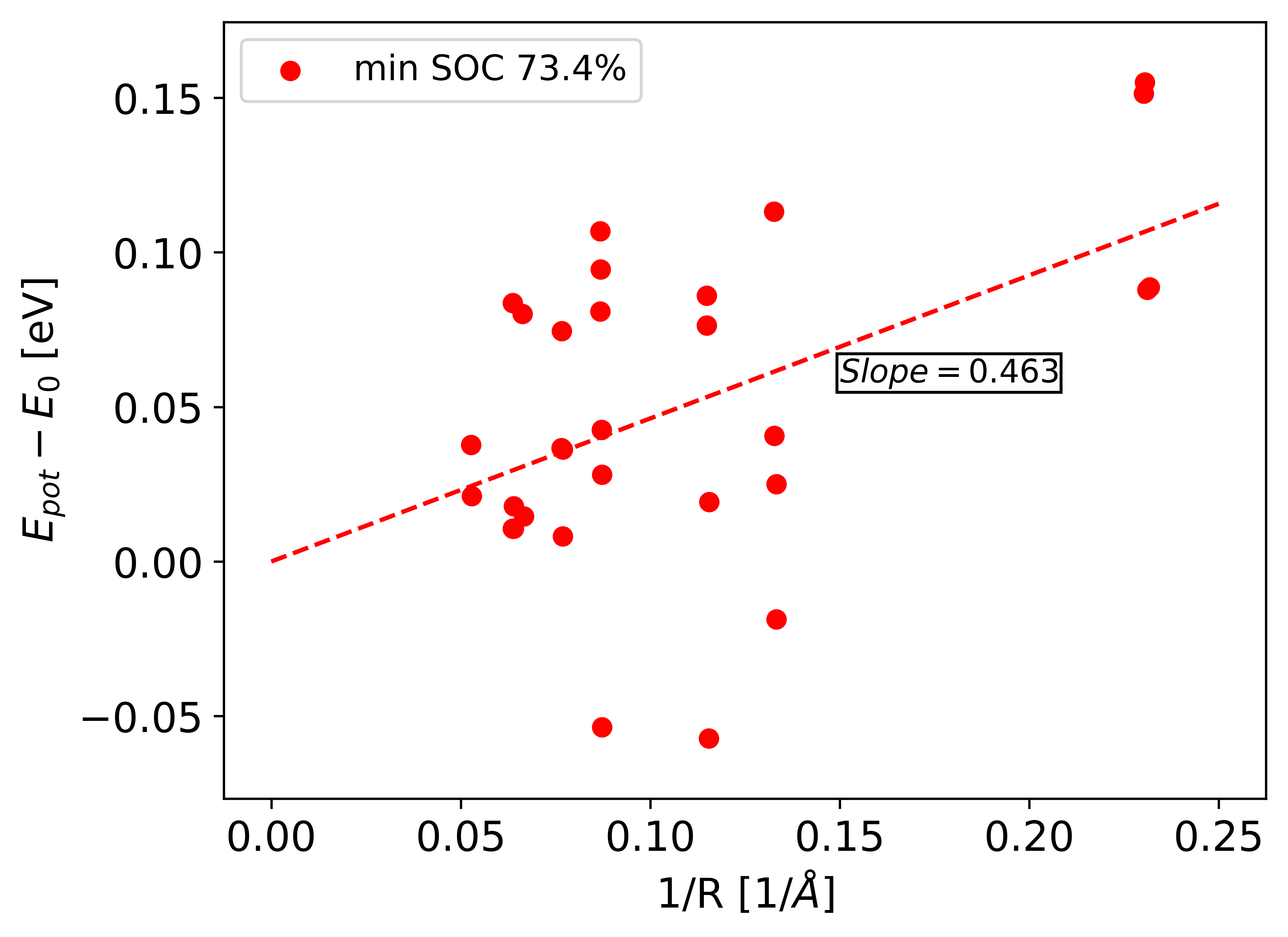}
 \caption{Approximate Coulomb energies $E_{Coul} \approx E_{pot}- E_0$ of local minima relative to $E_0$ (at the respective SOC) in dependence of the inverse distance between the investigated sites in the $inv(Li) - dilute$ (\textit{left}) and $inv(Vac) - dilute$ (\textit{right}) configurations. The global minima $E_0$ in the $1/R \rightarrow 0$ limit are extrapolated from the linear regression of the slope and set to $0$ in the plot.}
\label{fig:results_dilute}
\end{figure*}

 \begin{table*}[t] 
 \centering
 \begin{tabular}{lllll}
 \toprule
 \textbf{Structure} & \textbf{Stoichiometry} & \textbf{SOC} & \textbf{Slope [eV\text{\r{A}}]} & \textbf{rel. permittivity}\\
 \midrule
 $inv(Li) - dilute$ & \ce{Li_{32}C_{720}} & 26.7\% & $0.594\pm 0.078$ & $14.03\pm 3.12$ \\ 
 $inv(Vac) - dilute$ & \ce{Li_{94}C_{768}} & 73.4\% & $0.463\pm 0.152$ & $21.45\pm 9.52$ \\
 \bottomrule
 \end{tabular}
 \caption{Results for the data points constructed with dilutely filled adjacent layers. The slopes and their RMSEs stem from linear regression of the data points (minima). The corresponding relative permittivities have been calculated via the Coulomb law in~\cref{Diel:eq:4}.}
 \label{tab_dilute}
 \end{table*}
 
As expected, this dilutely filled adjacent layer causes a larger scatter of the potential energies of the individual configurations, which is due to the less even electron density of such a layer compared to a completely empty or full one. In the $inv(Vac) - dilute$ case, this effect is also combined with the already larger scatter we observed and explained previously when sampling this type of investigated layer. We tried to mitigate this by also sampling over two different orientations of the dilute adjacent layers -- averaging out the uneven electron density -- but the scatter remains very large, which leads to a much larger uncertainty of the resulting slope and relative permittivity $\epsilon_r$ (see~\cref{tab_dilute}).

Nevertheless, the Coulombic nature of the interaction is still clearly visible despite the scatter at least for the $inv(Li) - dilute$ case (\cref{fig:results_dilute}, left), and is expected to be captured by the (admittedly rough) linear fit in the other case as well -- yet of course with a larger uncertainty. Indeed, the resulting slope is still in line with the overall behaviour we find throughout the whole system.
 
\subsection{Dielectric screening in the $xy$-plane vs. in $z$-direction:}
\label{sub:z_dir}

Thus far, we have focused on the dielectric behaviour of Li-GICs parallel to the graphene sheets ($xy$-plane), because that is what governs the diffusion and intercalation of the Li-ions. However, real graphite anodes and experimental samples often come in powder form, with graphite nanoparticles in random spatial orientation. Experimental measurements of the dielectric behaviour of such samples cannot be directly compared to the results we have presented so far. Therefore, we also investigate the dielectric behaviour in $z$-direction (perpendicular to the graphene sheets).

 \begin{figure*}[h]
\includegraphics[width=0.99\columnwidth]{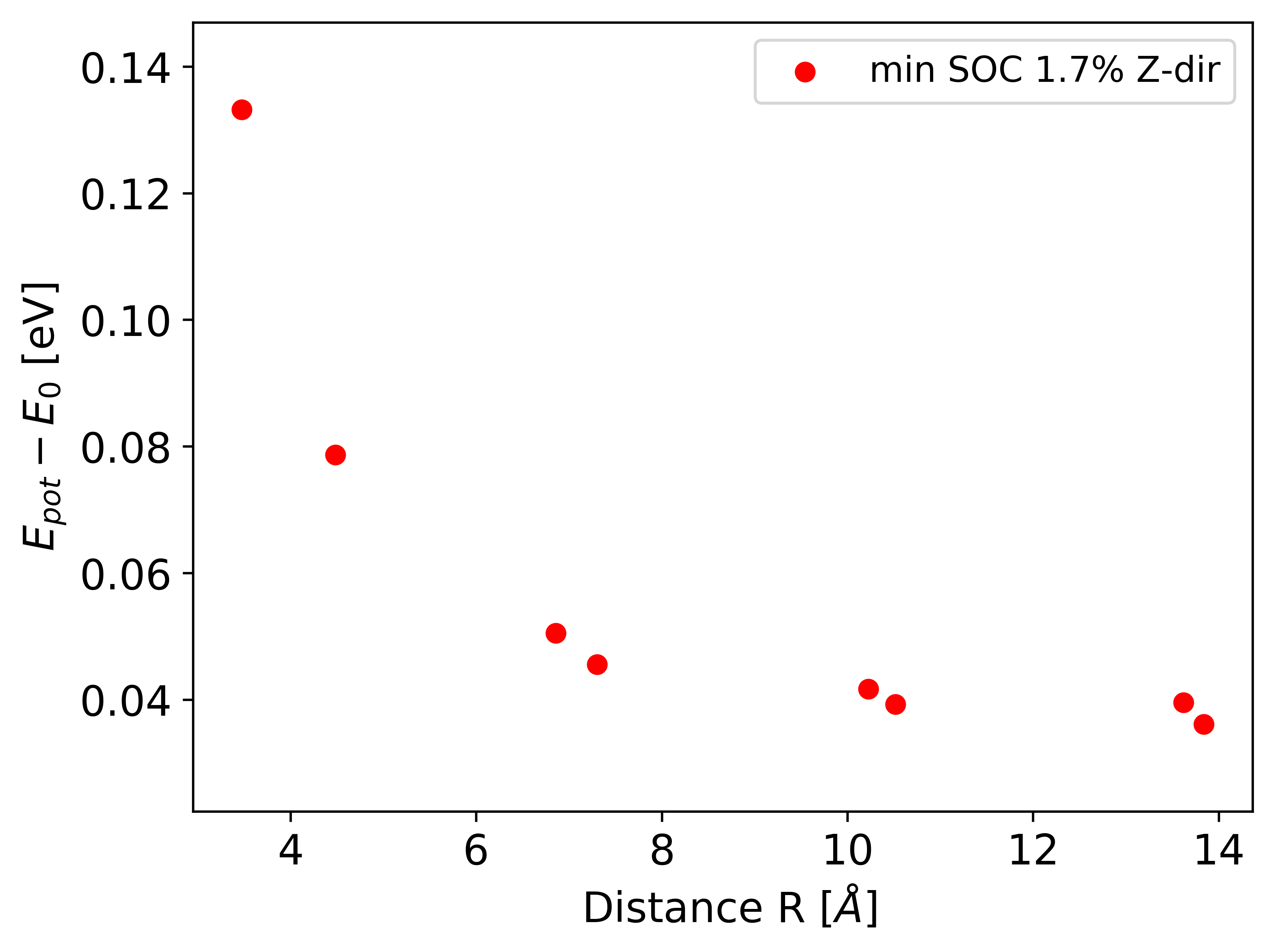}
\includegraphics[width=0.99\columnwidth]{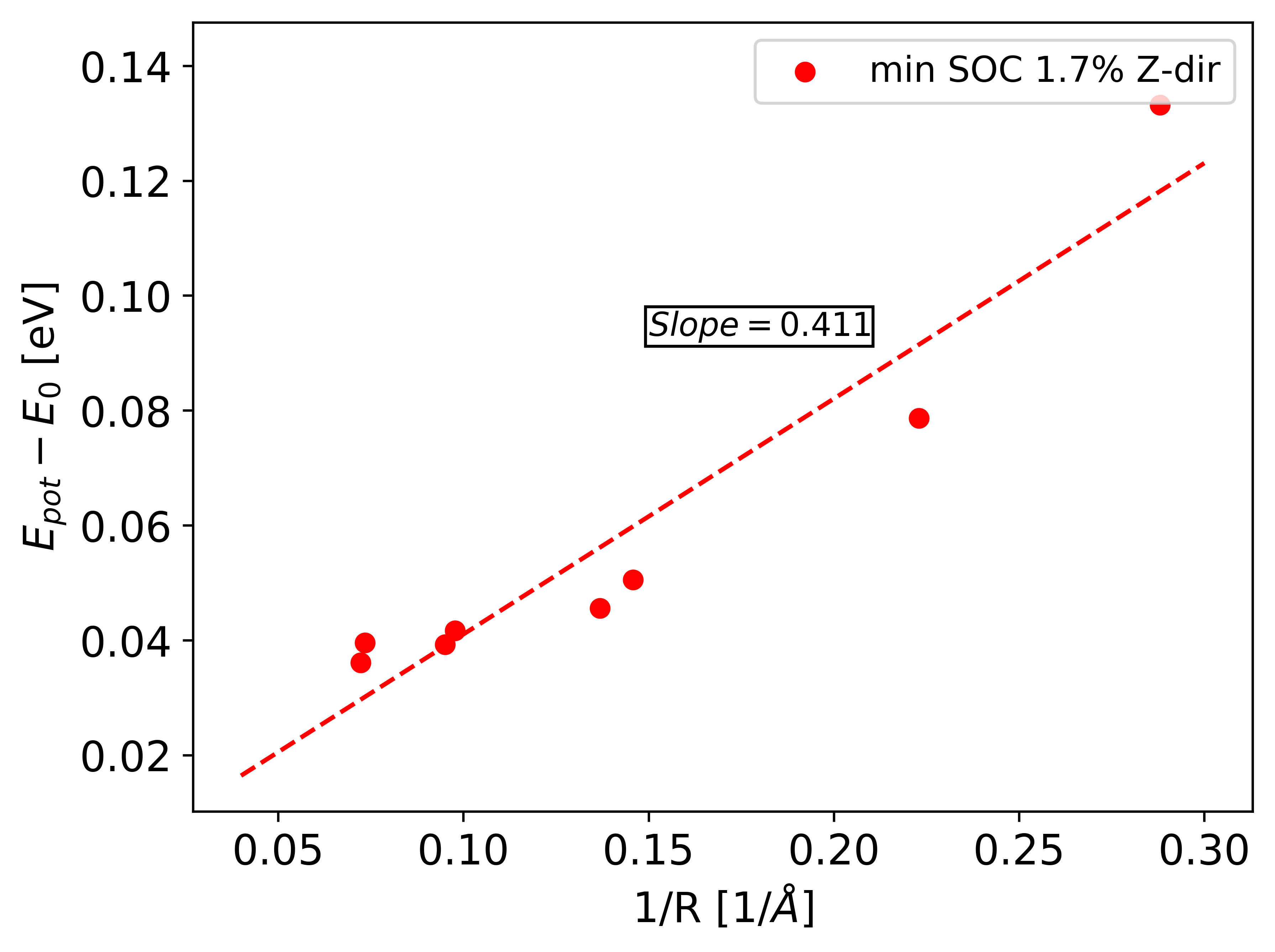}
 \caption{Approximate Coulomb energies $E_{Coul} \approx E_{pot}- E_0$ of local minima relative to $E_0$ (at the respective SOC) in dependence of the distance (\textit{left}) and inverse distance (\textit{right}) between the investigated Li-ions in $z$-direction.}
\label{fig:results_Z}
\end{figure*}

In order to do so, we construct a 15-layered supercell with a stoichiometry of \ce{Li2C_{720}} which corresponds to an SOC of 1.7\%. We sample 5 of the 15 layers (to ensure no self-interaction with the periodic image) with one pair of Li-ions, placed in positions both directly above each other and shifted by one next-neighbour position (those latter interactions are not perfectly in $z$-direction, but still ``through'' the graphene sheets). As can be seen, the resulting electrostatic behaviour is not purely Coulombic (see~\cref{fig:results_Z}).

Similar to the results for the investigated layer $inv(Vac)$, the presence of charge carriers in between the investigated Li positions seems to cause some extra, nonlinear screening. Holding on to the assumption of a linear trend, a linear fit finds a slope of $0.411\pm 0.054$, which translates to a relative permittivity of $\epsilon_z = 21.35\pm 5.50$ -- significantly larger than the in-plane relative permittivity $\epsilon_{xy}$ we find at similar SOC.
Based on this, we can average over the 3 spatial dimensions in order to provide a rough estimate of the relative permittivity $\epsilon_r$ of randomly oriented graphite powder at very low SOC, simply by applying:
 \begin{align} \epsilon_r = (2\epsilon_{xy}+\epsilon_z)/3 = 11.93 \label{Diel:eq:Z} \end{align} 
 This result is comparable to the $\epsilon_r \approx 15$ found by Hotta et al.~\cite{hotta2011complex} in the GHz range, but again, it is not entirely clear that this comparison is physically valid. Once a more reliable consensus is reached in experiment, our estimation of the partial charge can be further validated. It is also necessary to point out that grain-boundary and grain-size effects may play a significant role, especially the smaller the particles in the powder get, but these effects are not considered at all in the numbers we present here -- those are simply intended to help with comparison between our results and experiments.

\subsection{Relative permittivity $\epsilon_r$ as a function of the SOC:}
\label{sub:eps_soc}

\begin{figure*}[h]
 \centering
 \includegraphics[width=1.8\columnwidth]{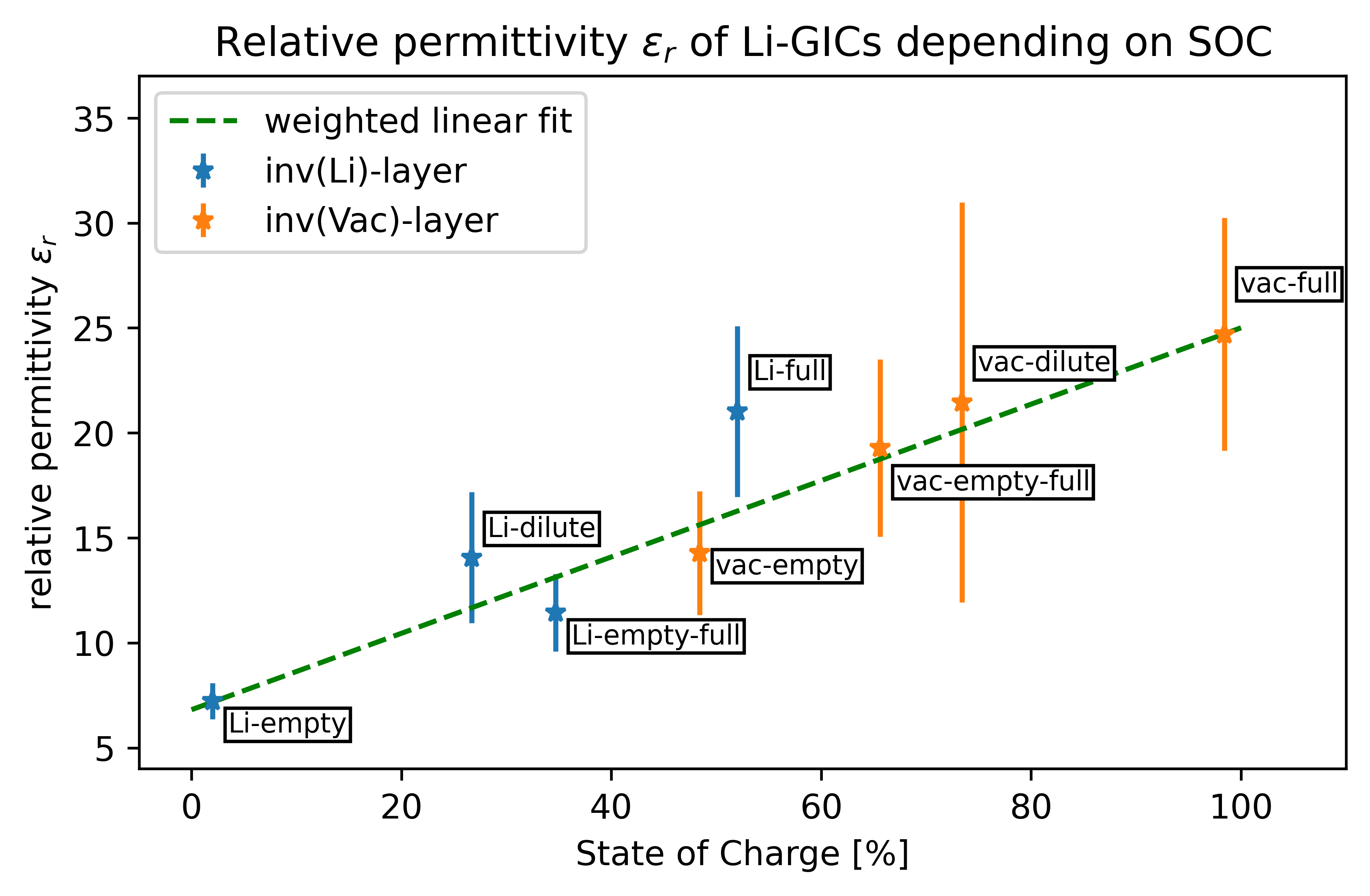}
 \caption{Relative permittivity $\epsilon_r$ in $xy$-plane (parallel to the graphene sheets), found for different configurations of the Li-GIC material as a function of the respective SOC . The weighted linear fit outputs the expression $\epsilon_r(SOC) = 6.8 + 18.2\cdot SOC$.}
 \label{fig:results}
\end{figure*}

Putting the previous results together (see~\cref{fig:results}), we find an approximately linear dependency of the relative permittivity $\epsilon_r$ (in the $xy$-plane) on the SOC of the Li-GIC. Between 20\% and 80\% SOC, we observe some additional effects of the local ordering -- dilute or staged -- and of whether Li-ions in the otherwise empty layers or Li-ion vacancies in the otherwise full layers are sampled. These deviations are largest at around 50\% SOC, but the average $\epsilon_r$ of configurations at roughly equal SOC are still close to the weighted linear regression fit, so the more macroscopic a viewpoint is taken, the less these local phenomena matter.
 
The final uncertainty intervals are translated through~\cref{Diel:eq:4} and stem from a combination of our (arbitrarily chosen) uncertainty in the partial charge $Z_{Li}$ (as described in the methodology section) and the root mean square error (RMSE) of the slope, which is returned by the linear regression and caused by scatter around the Coulombic behaviour, which in turn is due to variations in the electron densities of adjacent layers, local structural perturbations and similar effects not captured by the approximations we made when introducing $E_{Coul} \approx E_{pot}- E_0$ (as described throughout the previous sections), as well as the intrinsic limited accuracy of DFTB.

\section {CONCLUSION:}
\label{sec:conclusion}

With this work, we present the first rigorous investigation of the dielectric behaviour of lithium graphite intercalation compounds (Li-GICs) for the entire functional range of charge during the application as anode of a modern Li-ion battery. In doing so, we provide a straightforward approach for investigating the intrinsic relative permittivity of materials with mobile charge carriers in the electrostatic limit, that can be applied to other materials in the future -- given that sufficiently fast and long-ranged computational methods are available. Thanks to our recently published DFTB parametrization~\cite{panosetti2021dftb, annies2021accessing}, we are able to sample the long-ranged Coulomb interactions between two intercalated charge carriers (Li-ions or vacancies) in a variety of stoichiometries and configurational realizations thereof. By examining the approximations we made during this process, we additionally outline ways to further improve this methodology in the future.

The primary finding of this work is the mostly linear dependency (from $\epsilon_r \approx 7$ at SOC 0\% to $\epsilon_r \approx 25$ at SOC 100\%) of the relative permittivity on the state of charge, which we put forward for the first time. With this, we make valuable contributions to the future modeling of functional materials by means of charged kinetic Monte Carlo and continuum simulations and to the general understanding of Li-GICs. Our results hold for qualitatively different realizations of the intermediate stoichiometries (staged or dilute), as well as both possible types of diffusion (Li-ion or vacancy). The few available experimental studies agree reasonably well with our results, but more investigation is necessary to really pinpoint the quantitative dielectric response, especially at higher SOC. Thanks to our results, it is now clear that only two measurements -- one at low and one at high SOC -- would suffice for that purpose.

We find that an approximation neglecting the local distortions of the structure and of the electron density caused by the sampled charge carriers holds very well if those charge carriers are Li-ions, and slightly less well, but still within reason, in the vacancy case. Future improvements could be achieved by correcting with some local descriptor based machine learning model that is specifically trained to pick up those local effects.

Additionally, we provide a rough estimate (grain-boundary effects are neglected, but should be uniform with space direction and therefore should not have any qualitative impact on our results) on how the relative permittivity of a perfect crystal and a powder can be compared, or relatedly, of how an experimental result for (intercalated) graphite powder can be translated to the internal $xy$-plane relative permittivity the Li-ions feel locally (which is the one that is actually relevant for their diffusion behaviour).

\section{Declaration of competing interests}
The authors declare that they have no known competing financial interests or personal relationships that could have appeared to influence the work reported in this paper.

\section{Acknowledgements}
This work was funded by the German Federal Ministry of Education and Research (BMBF) as part of the research cluster ``AQua'' within the project InOPlaBat (grant number 03XP0352). The authors gratefully acknowledge the computational and data resources provided by the Leibniz Supercomputing Centre (LRZ). The authors jointly thank Sebastian Matera, Jakob Filser, David Egger, Cristina Grosu and Julian Holland for fruitful discussions.

\bibliographystyle{elsarticle-num-names} 
\bibliography{bibliography}
\end{document}